\documentclass[11pt]{article}
\usepackage{amssymb}
\usepackage{amsmath}
\usepackage{amsthm}
\usepackage{graphicx}
\newtheorem{theorem}{Theorem}[section]
\newtheorem{corollary}[theorem]{Corollary}
\newtheorem{lemma}[theorem]{Lemma}

\newtheorem{property}{Property}

\newtheorem{conjecture}{Conjecture}

\newtheorem{definition}[theorem]{Definition}
\newtheorem{question}[conjecture]{Question}

\def\OPT#1#2#3{\mbox{OPT}_{#1,#2}\left(#3\right)}

\def\vs#1#2#3{#1_{#2},\dots,#1_{#3}}

\def\setof#1{\left\{{\let\st\colon #1 }\right\}}

\def\Reals#1{\mathbb{R}^{#1}}
\def\pnorm#1#2{\left\| #2 \right\|_{#1}}

\def\orig#1{\bar{#1}}
\def\form#1#2{\left\langle#1 | #2 \right\rangle}

\def\allone{\vec{\mathbf{1}}}

\def\aa{\mathbf{a}}
\def\bb{\mathbf{b}}
\def\dd{\mathbf{d}}
\def\ee{\mathbf{e}}
\def\pp{\mathbf{p}}
\def\uu{\mathbf{u}}
\def\ww{\mathbf{w}}
\def\xx{\mathbf{x}}
\def\yy{\mathbf{y}}

\def\DDelta{\mathbf{\Delta}}
\def\AA{\mathbf{A}}
\def\BB{\mathbf{B}}
\def\DD{\mathbf{D}}
\def\EE{\mathbf{E}}
\def\II{\mathbf{I}}

\def\XX{\mathbf{X}}

\def\symP{\mathbb{P}}

\def\smoothed#1#2{\mbox{Smoothed}_{#1}\left[#2 \right]}
\def\diag#1{\mbox{diag}\left(#1\right)}
\def\expec#1#2{\mbox{\rm E}_{#1}\left[ #2 \right]}
\newcommand{\set}[1]{\left\{#1\right\}}

\def\pp{\mathbf{p}}
\def\qq{\mathbf{q}}
\def\ss{\mathbf{s}}
\def\ttt{\mathbf{t}}
\def\uu{\mathbf{u}}
\def\ww{\mathbf{w}}
\def\zz{\mathbf{z}}

\def\XX{\mathbf{X}}

\def\NN{\mathbf{N}}
\def\ZZ{\mathbf{Z}}

\def\calN{\mathcal{N}}
\def\calM{\mathcal{M}}

\def\allzero{\vec{\mathbf{0}}}

\begin{document}

\title{On the Approximation and Smoothed Complexity of \\
       Leontief Market Equilibria}

\author{
Li-Sha Huang \\
Department of Computer Science\\
Tsinghua University\\
Beijing, China
\and 
Shang-Hua Teng 
\thanks{Also affiliated with Akamai Technologies Inc. Cambridge, 
 Massachusetts, USA.
Partially supported by NSF grants CCR-0311430 and 
  ITR CCR-0325630. Part of this work done while visiting
Tsinghua University.}\\
Computer Science Department\\
Boston University\\
Boston, Massachusetts, USA
}
\date{}

\maketitle

\begin{abstract}
We show that the problem of finding an
   $\epsilon$-approximate Nash equilibrium
   of an $n\times n$ two-person games
    can be reduced to
   the computation of an $(\epsilon/n)^2$-approximate market
   equilibrium of a Leontief economy.
Together with a recent result of Chen, Deng and Teng, this
  polynomial reduction implies that
  the Leontief market exchange problem
    does not have a fully polynomial-time
  approximation scheme, that is, there is no algorithm that
  can compute an $\epsilon$-approximate market equilibrium
  in time polynomial in $m$, $n$, and $1/\epsilon$,
  unless \textbf{PPAD} $\subseteq$ \textbf{P},
 We also extend the analysis of
   our reduction to show,
   unless  \textbf{PPAD} $\subseteq$ \textbf{RP},
   that the smoothed complexity of the Scarf's general fixed-point
   approximation algorithm (when applying to solve
   the approximate Leontief market exchange problem) or of any algorithm for
   computing an approximate market equilibrium
   of Leontief economies is not polynomial
   in $n$ and $1/\sigma$,
   under Gaussian or uniform perturbations with magnitude $\sigma$.
\end{abstract}

\newpage

\section{Introduction}

A {\em Leontief economy} \cite{ScarfBook,YeRationality} with $m$ divisible
  goods (or commodities) and $n$ traders
  is specified by a pair of  non-negative $m\times n$
   matrices\footnote{See Section \ref{sec:notations} for the
   basic notations used in this paper.}
  $\EE = (e_{i,j})$ and $\DD = (d_{i,j})$.
After trading, the exchange of goods can also
  be expressed by a non-negative $m\times n$ matrix $\XX = (x_{i,j})$.

\begin{itemize}
\item
The matrix $\EE$ is the {\em endowment matrix}
   of the traders, that is, $e_{i,j}$ is the amount of
   commodity $i$ that trader $j$ initially has.
We can assume, without loss of generality, that there is
  one unit of each type of good.
With this assumption, the sum of every row of $\EE$ is equal to 1.

\item The matrix $\DD$ is the {\em demand matrix} or the
  {\em utility matrix}.
It defines $n$ utility functions $\vs{u}{1}{n}$, one for each trader.
For each exchange $\XX$ of goods, let $\xx_j$ be the
  $j^{th}$ column of $\XX$, then
\[
u_j(\xx_j) = \min_i \left\{  \frac{x_{i,j}}{d_{i,j}}\right\}
\]
is the {\em Leontief utility} of trader $j$.
\end{itemize}

The initial utility of trader $j$ is $u_j(\ee_j)$ where $\ee_{j}$ is
  the $j^{th}$ column of $\EE$.
The {\em individual objective} of each trader is to maximize his or her
  utility.
However, the utilities that these traders can achieve depend on
  the initial endowments, the individual utilities, and
  potentially, the (complex) process that they perform their
  exchanges.

In Walras' pure view of economics \cite{WalrasFranch},
  the individual objectives of traders and their initial
  endowments enable the market to establish a price vector $\pp$ of
  the goods in the market.
Then the whole exchange can be conceptually characterized as:
   the traders sell their endowments -- to obtain money or budgets -- and
   individually  optimize their objectives by buying the bundles of goods
   that maximize their utilities.

By selling the initial endowment $\ee_j$, trader $j$
  obtains a budget of $\form{\ee_j}{\pp}$ amount, where
  $\form{\ee_j}{\pp}$ denotes the dot-product of these two
  vectors.
The {\em optimal bundle} for $u_j$ is a solution to the
  following mathematical program:
\begin{eqnarray}\label{eqn:individual}
\max u_j(\xx_j) \quad \mbox{ subject to } \form{\xx_j}{\pp} \leq
   \form{\ee_j}{\pp}.
\end{eqnarray}

A solution to Equation (\ref{eqn:individual}) is referred to
  as an {\em optimal demand} of trader $j$ under prices $\pp$.
  The price vector $\pp$ is a {\em Walrasian equilibrium},
  an {\em Arrow-Debreu equilibrium}, or simply an {\em equilibrium}
  of the Leontief
  economy $(\EE,\DD)$ if
  there exists optimal solution
  $\XX = \left(\vs{\xx}{1}{n}\right)$ to
  Equation (\ref{eqn:individual}) such that
\begin{eqnarray}\label{eqn:Walrasian}
\sum_{j} \xx_j \leq \allone,
\end{eqnarray}
where $\allone$ is the $m$-dimensional column vector with all ones.
This last constraint states that the traders' optimal demands can be
  met by the market.

In other words, an equilibrium price vector essentially
  allows each trader to make individual decision without considering
  others' utilities nor how they achieve their objectives.

The computation of a market equilibrium is a fundamental problem
  in modern economics \cite{ScarfBook} as Walrasian equilibria might provide
  useful information for the prediction of market trends,
  in the decision for future investments,
  and in the development of economic policies.
So a central complexity question in Leontief market exchange problem is:

\begin{question}[Polynomial Leontief?]\label{que:poly}
Is the problem of computing an equilibrium
  of a Leontief economy in \emph{\textbf{P}}?
\end{question}

So far no polynomial-time algorithm has been found for this problem.
In practice, one may be willing
  to relax the condition of equilibria
  and considers the computation of
  approximate market equilibria.
For example, Scarf \cite{ScarfFixedPointApproximation}
  developed a general algorithm for computing approximate
  fixed points and equilibria.
Recently, Deng, Papadimitriou, Safra \cite{DPS}
  proposed a notion of an approximate market equilibrium as
  a price vector that allows each trader to independently
  and approximately optimize her utilities.

For any $\epsilon \geq 0$, let $\OPT{j}{\epsilon}{\pp,\ee_j}$
  be the set of $\epsilon$-approximately optimal vectors in $\Reals{m}_+$
  for Equation (\ref{eqn:individual}), that is, the set of all $\xx_j$
  satisfying
\begin{eqnarray*}
  \form{\xx_j}{\pp} & \leq &   (1+\epsilon)\form{\ee_j}{\pp}, \quad \mbox{ and } \\
   u_j(\xx_j) & \geq &
  (1-\epsilon)u(\xx_j'),\quad \forall \xx_j': \form{\xx_j'}{\pp} \leq
  \form{\ee_j}{\pp}
\end{eqnarray*}

Then $\pp$ is an {\em $\epsilon$-approximate
  equilibrium}\footnote{One can of course define a stronger notion of
approximate equilibria: $\pp$ is an {\em $\epsilon$-strictly approximate
  equilibrium} if there exists $\XX = (\vs{\xx}{1}{n})$ such that
\begin{eqnarray*}
\form{\xx_j}{\pp} & \leq &   \form{\ee_j}{\pp}, \\
   u_j(\xx_j) & \geq &
  (1-\epsilon)u(\xx_j'),\quad \forall \xx_j': \form{\xx_j'}{\pp} \leq
  \form{\ee_j}{\pp}, \quad { and }\\
\sum_{j} \xx_j & \leq & \allone.
\end{eqnarray*}
It is easy to show that every $(\epsilon/2)$-approximate Walrasian
  equilibrium $\pp $ is an $\epsilon$-strictly approximate Walrasian
  equilibrium, by dividing its associated exchange $\XX$ by a
factor of $(1-\epsilon /2)$.
Therefore, in the remainder of this paper, we will,
  without loss of generality, stay with the
  less restrictive notion of approximation equilibria.
}
  of a Leontief economy $(\EE,\DD)$
  if  there exists $\epsilon$-approximately optimal solution
  $\XX = \left(\vs{\xx}{1}{n}\right)$
  with $\xx_j \in \OPT{j}{\epsilon}{\pp,\ee_j}$
   such that
\begin{eqnarray}\label{eqn:approxWalrasian}
\sum_{j} \xx_j \leq (1+\epsilon)\allone.
\end{eqnarray}

\begin{question}[Fully Polynomial Approximate Leontief?]
\label{que:approxPoly}
Can an $\epsilon$-approximate equilibrium
  of a Leontief economy with $m$ goods and $n$ traders be computed in
  time polynomial in $m$, $n$, and $1/\epsilon$?
\end{question}

The combination of two recent results greatly dashed the hope
  for a positive answer to Question \ref{que:poly}.
Codenotti, Saberi, Varadarajan, and Ye \cite{CSVY}
  gave a polynomial-time reduction from two-person games
  to a special case of the Leontief economy.
In a remarkable breakthrough,
  Chen and Deng \cite{ChenDeng} subsequently proved that
  the problem of finding a Nash equilibrium of a
  two-person game is \textbf{PPAD}-complete\footnote{We refer 
   the readers who are not familiar with the complexity class
  \textbf{PPAD} to the paper by Papadimitriou \cite{PAP94}
  and also to the subsequent papers on the \textbf{PPAD}-completeness
  of normal games \cite{DAS05,CHE05,DAS06,ChenDeng,ChenDengTeng}.
This class includes several important search problems such as
  various discrete fixed-point problems and the problem of
  finding a Nash equilibrium of $r$-person games for any fixed $r$.
}.

In this paper, we show that there is no fully polynomial-time
  approximation scheme for Leontief economies
  unless \textbf{PPAD} $\subseteq$ \textbf{P}.
Hence it is unlikely that Question \ref{que:approxPoly}
  has a positive answer.
By analyzing the numerical properties of the reduction of
  Condenotti, Saberi, Varadarajan, and Ye \cite{CSVY},
  we prove that the problem of finding an $\epsilon$-approximate
  Nash equilibrium of an $n\times n$ two-person games can be reduced to
  the computation of an $(\epsilon/n)^2$-approximate
   equilibrium of a Leontief economy with $2n$ goods and $2n$ traders.
This polynomial relationship between
  approximate Nash equilibria of bimatrix games and
  approximate Walrasian equilibria of Leontief economies
  is significant because it enables us to apply the recent
  result of Chen, Deng, and Teng \cite{ChenDengTeng}
  to show that finding an approximate market equilibrium
  with only $O(\log n)$-bits of precision is as hard as finding an
  exact market equilibrium, which in turn is as hard as
  finding a Nash equilibrium of a two-person game
  or a discrete Brouwer fixed point in the most general settings.

We also consider the smoothed complexity of the Leontief market exchange
  problem. In
  the smoothed model introduced by Spielman and Teng
  \cite{SpielmanTengSimplex}, an algorithm receives and solves a perturbed
  instances.
The smoothed complexity of an algorithm is the maximum
  over its inputs of the expected running time of the algorithm under
  slight random perturbations of that input.
The smoothed complexity is
  then measured as a function of both the input length and the
  magnitude $\sigma$ of the perturbations.
An algorithm has {\em smoothed polynomial-time complexity}
  if its smoothed measure
  is polynomial in $n$, the problem size, and in $1/\sigma$
  \cite{SpielmanTengSimplex,SpielmanTengSurvey}.

In the smoothed model for Leontief economies,
  we start with a pair of $m\times n$ matrices
  $\orig{\EE} = \left( \orig{e}_{i,j}\right)$
  and $\orig{\DD} = \left(\orig{d}_{i,j}\right)$
  with $0\leq \orig{d}_{i,j} \leq 2$
  and $0\leq \orig{e}_{i,j} \leq 1$.
Suppose $\EE = \left( e_{i,j}\right)$ and $\DD = \left(
 d_{i,j}\right)$ are perturbations of $\orig{\EE}$ and $\orig{\DD}$ where
  $e_{i,j} = \max(0,\orig{e}_{i,j} + r^{E}_{i,j})$ and
   $d_{i,j} = \max(0,\orig{d}_{i,j} + r^{D}_{i,j})$,
    with $r^{E}_{i,j}$ and $r^{D}_{i,j}$ being chosen
    independently and uniformly from $[-\sigma,\sigma]$.
The {\em smoothed complexity} of the Leontief exchange problem
  $(\orig{\EE},\orig{\DD})$ is then
  measured by the expected complexity of finding an
  equilibrium of the Leontief economy $(\EE,\DD)$.

The following has been an open question in the smoothed analysis of
  algorithms.
\begin{question}[Smoothed Polynomial Leontief?]
Can an equilibrium of a Leontief economy be computed
  in smoothed time polynomial
  in $m$, $n$, and $1/\sigma$?

Can an $\epsilon$-equilibrium of a
    Leontief economy be computed
  in smoothed time polynomial
  in $m$, $n$, $1/\epsilon$ and $1/\sigma$?
\end{question}

A concrete open question has been whether the smoothed complexity of
  the classic Scarf's general fixed-point approximation
  algorithm \cite{ScarfFixedPointApproximation}  is polynomial
   for solving the Leontief market exchange problem.

By refining our analysis of the reduction from the two-person games
to Leontief economies, we show it is unlikely that Scarf's algorithm
  has polynomial smoothed complexity for computing an approximate
  equilibrium of Leontief economies.
In particular, we prove that, unless \textbf{PPAD} $\subseteq$ \textbf{RP},
the problem of finding an (approximate)
  equilibrium of a Leontief economy is
   not in smoothed polynomial time.

\section{Notations}\label{sec:notations}

We will use bold lower-case
  Roman letters such as $\xx $, $\aa$, $\bb_{j}$ to denote vectors.
Whenever a vector, say $\aa\in\Reals{n}$ is
  present, its components will be denoted by
  lower-case Roman letters with subscripts, such as $\vs{a}{1}{n}$.
Matrices are denoted by bold upper-case Roman letters
  such as $\AA$ and scalars are usually denoted by lower-case Roman
  letters.
The $(i,j)^{th}$ entry of a matrix $\AA$ is
  denoted by $a_{i,j}$.
We use $\aa_i$ to denote the $i^{th}$
  column of $\AA$.

We now enumerate some other notations that are used in this paper.

\begin{itemize}
\item $\Reals{m}_+$:  the set of $m$-dimensional vectors with non-negative
  real entries.

\item $\Reals{m\times n}_{[a:b]}$: the set of all $m \times n$ matrices
      with real entries between $a$ and $b$.
For example, $\Reals{m\times n}_{[0:2]}$ is
   the set of  non-negative matrices
   with entries at most 2.

\item $\symP^n$: the set of all vectors $\xx$ in $n$ dimensions
  such that $\sum_{i=1}^n x_i = 1$ and $x_i\geq 0$ for all $1\leq i\leq n$.
\item $\form{\aa}{\bb}$: the dot-product of two vectors in the same dimension.

\item $\pnorm{p}{\xx}$: the $p$-norm of vector $\xx$, that is,
$\left(\sum |x_i^p|\right)^{1/p}$ and 
$\pnorm{\infty}{\xx} = \max_i |x_i|$.
\end{itemize}

\section{Approximate Nash Equilibria of Two-Person Games}

The non-zero-sum {\em two-person game} or the {\em bimatrix game}
   is a non-cooperative game between two players
   \cite{NashNonCooperative,Lemke,LemkeHowson}, the
   {\em row player} and the {\em column player}.
If  the row player has $m$ pure strategies and
  the column player has $n$ pure strategies,
then their payoffs are given by a pair of $m\times n$ matrices
  $(\AA,\BB)$.

A mixed row strategy is a vector $\xx \in \symP^m$ and
  a mixed column strategy is a vector $\yy \in \symP^n$.
The expected payoffs to these two players are respectively
  $\xx^{\top} \AA \yy$ and $\xx^{\top} \BB \yy$.
A {\em  Nash equilibrium} is then a pair of vectors
  $(\xx^{*}\in \symP^m,\yy^{*}\in \symP^n)$ such that for all pairs of vectors
  $\xx \in \symP^m$ and $\yy \in \symP^n$,
\[
(\xx^{*})^{\top} \AA  \yy^{*} \geq \xx^{\top} \AA  \yy^{*} \quad
\mbox{and} \quad (\xx^{*})^{\top} \BB \yy^{*}  \geq
(\xx^{*})^{\top} \BB \yy.
\]
Every two-person game has at least one Nash
  equilibrium \cite{NashNonCooperative}.
But in a recent breakthrough, Chen and Deng \cite{ChenDeng}
  proved that the problem of computing a Nash equilibrium of a
  two-person game is \textbf{PPAD}-complete.

One can relax the condition of Nash equilibria and considers
  approximate Nash equilibria.
There are two possible notions of approximation.

\begin{definition}[Approximate Nash equilibria]
An {\em $\epsilon$-approximate Nash equilibrium} of game $(\AA,\BB)$
is a pair of mixed strategies $(\xx^*,\yy^*)$, such that for
all $\xx,\yy\in \symP^n$,
\begin{equation*}
(\xx^{*})^{\top} \AA  \yy^{*} \geq \xx^{\top} \AA
\yy^{*}-\epsilon \quad \mbox{and} \quad (\xx^{*})^{\top} \BB
\yy^{*}  \geq (\xx^{*})^{\top} \BB \yy-\epsilon.
\end{equation*}
\end{definition}

\begin{definition}[Relatively Approximate Nash equilibria]
An {\em  $\epsilon$-relatively-approximate Nash
  equilibrium} of game $(\AA,\BB)$
  is a pair of mixed strategies $(\xx^*,\yy^*)$, such that for
  all $\xx,\yy\in \symP^n$,
\begin{equation*}
(\xx^{*})^{\top} \AA  \yy^{*} \geq (1-\epsilon) \xx^{\top} \AA
\yy^{*} \quad \mbox{and} \quad (\xx^{*})^{\top} \BB \yy^{*}  \geq
(1-\epsilon) (\xx^{*})^{\top} \BB \yy.
\end{equation*}
\end{definition}

Note that the Nash equilibria and the relatively-approximate
  Nash equilibria of a  two-person  game
  $(\AA,\BB)$ are invariant under positive scalings,
  i.e., the bimatrix game $(c_1\AA,c_2\BB)$
  has the same set of Nash equilibria  and relatively-approximate
  Nash equilibria as the bimatrix game
  $(\AA,\BB)$, as long as $c_1, c_2>0$.
However,  each $\epsilon$-approximate
    Nash equilibrium $(\xx,\yy)$ of $(\AA,\BB)$
   becomes
    a $c\cdot\epsilon$-approximate Nash equilibrium of the bimatrix game
    $(c\AA,c\BB)$ for $c>0$.

On the other hand, Nash equilibria and approximate Nash equilibria are
  invariant under shifting, that is,
  for any constants $c_1$ and $c_2$,
  the bimatrix game $(c_1+\AA,c_2+\BB)$
  has the same set of Nash equilibria and approximate Nash
  equilibria as the bimatrix game $(\AA,\BB)$.
However, shifting may not preserve the relatively-approximate
  Nash equilibria.

Thus, we often normalize the matrices $\AA$ and $\BB$
  so that all their entries are between 0 and 1, or between
  -1 and 1, in order to study the complexity of approximate Nash
  equilibria \cite{Lipton,ChenDengTeng}.

Recently, Chen, Deng, and Teng \cite{ChenDengTeng} proved the
  following result.

\begin{theorem}[Chen-Deng-Teng]
The problem of computing an $1/n^6$-approximate
  Nash equilibrium of a normalized $n\times n$ two-person game
  is \emph{\textbf{PPAD}}-complete.
\end{theorem}

As pointed out in \cite{ChenDengTeng}, the 6 in the exponent
  of the above theorem can be replaced by any positive constant.
One can easily derive the following corollary.
\begin{corollary}[Relative approximation is also hard]\label{cor:cor}
It remains \emph{\textbf{PPAD}}-complete to compute a
  $1/n^{\Theta(1)}$-relatively-approximate
  Nash equilibrium of an $n\times n$ two-person game.
\end{corollary}

\section{Leontief Market Equilibria:
    Approximation and Smoothed Complexity}

In this and the next sections, we analyze a reduction $\pi$ that transforms
  a two-person game $(\orig{\AA},\orig{\BB})$ into a Leontief economy
  $(\orig{\EE},\orig{\DD}) = \pi(\orig{\AA},\orig{\BB})$
   such that from each $(\epsilon/n)^2$-approximate
  Walrasian equilibrium of $(\orig{\EE},\orig{\DD})$ we can construct
  an $\epsilon$-relatively-approximate Nash
  equilibrium of $(\orig{\AA},\orig{\BB})$.

We will also consider the smoothed complexity of Leontief economies.
To establish a hardness result for computing an (approximate) market
  equilibrium in the smoothed model,
  we will examine the relationship of Walrasian equilibria
  of a perturbed instance $(\EE,\DD)$ of $(\orig{\EE},\orig{\DD})$
  and approximate Nash equilibria of $(\orig{\AA},\orig{\BB})$.
In particular, we show that if the magnitude of the perturbation
  is $\sigma$, then we can construct an
  $(\epsilon+n^{1.5}\sqrt{\sigma})$-relatively-approximate
  Nash equilibrium of $(\orig{\AA},\orig{\BB})$
  from each $(\epsilon/n)^2$-approximate
   equilibria of $(\EE,\DD)$.

Because the analysis needed for approximate Walrasian equilibria
  is a special case (with $\sigma = 0$)
  of the analysis for the smoothed model,
  we will write only one proof for this general case,
  and present it in the next section.

In this section, we discuss the reduction from
  two-person games to Leontief economies, connect
  the smoothed model with approximation,
  and present the main theorems of this paper.

\subsection{Approximate Market Equilibria of Leontief Economy}


We first introduce a form of approximate market equilibria that
  is easier for the analysis of the reduction between game equilibria and
  market equilibria.

Let $\DD$ be the demand matrix and $\EE$ be the endowment matrix
  of a Leontief economy with $m$ goods and $n$ traders.
Given a price vector $\pp$, trader $j$ can obtain a budget of
  $\form{\ee_j}{\pp}$ by selling the endowment.
By a simple variational argument, one can show that the optimal
  demand $\xx_j$ with budget $\form{\ee_j}{\pp}$ satisfies
  $x_{i,j}/d_{i,j} = x_{i',j}/d_{i',j}$
  for all $i$ and $i'$ with $d_{i,j} > 0$ and $d_{i',j} > 0$.
Thus, under the price vector $\pp$, the maximum utility
  that trader $j$ can achieve is $0$ if $\form{\ee_j}{\pp} = 0$,
  and $\form{\ee_j}{\pp}/\form{\dd_j}{\pp}$ otherwise.
Moreover, in the latter case,
   $x_{i,j} = d_{i,j}\left(\form{\ee_j}{\pp}/\form{\dd_j}{\pp}\right)$.
Let $\uu = (\vs{u}{1}{n})$ denote the vector of utilities of the traders.
Then $\pp$ is a Walrasian equilibrium price if
\begin{eqnarray} \label{eqn:equilibriaMarket}
\pp  \geq  \allzero,\quad u_i=
\frac{\form{\ee_i}{\pp}}{{\form{\dd_i}{\pp}}}, \quad
\mbox{and}\quad \DD\uu \leq \allone.
\end{eqnarray}

In the remainder of this paper, we will refer to
  a pair of vectors $(\uu,\pp)$ that satisfies
   Equation (\ref{eqn:equilibriaMarket}) as an {\em equilibrium}
   of the Leontief economy $(\EE,\DD).$
Then, an {\em $\epsilon$-approximate equilibrium}
   of the Leontief economy $(\EE,\DD)$ is a pair of
   utility  and price vectors $(\uu,\pp)$ satisfying:
\[\left\{\begin{array}{rcll}
u_i&\geq&(1-\epsilon)\form{\ee_i}{\pp}/\form{\dd_i}{\pp},\forall i.\quad&\mbox{ --- All traders are approximately satisfied.}\\
u_i&\leq&(1+\epsilon)\form{\ee_i}{\pp}/\form{\dd_i}{\pp},\forall i.&\mbox{ --- Budget constraints approximately hold.}\\
\DD\uu&\leq&(1+\epsilon)\cdot\allone.&\mbox{ --- The demands
approximately meet the supply.}\\
\end{array}\right.\]

Note that if $(\uu,\pp)$ is an equilibrium of $(\EE,\DD)$, so
  is $(\uu,\alpha\pp)$ for every $\alpha > 0$.
Similarly, if $(\uu,\pp)$ is an $\epsilon$-equilibrium of $(\EE,\DD)$, so
  is $(\uu,\alpha\pp)$ for every $\alpha > 0$.
Thus, we can normalize $\pp$ so that $\pnorm{1}{\pp} = 1$.
In addition, for approximate equilibria, we assume
  without loss of generality that $\uu$ and $\pp$ are
  strictly positive to avoid division-by-zero
  since a small perturbation of an approximate equilibrium is
  still a good approximate equilibrium.

\subsection{Reduction from NASH to LEONTIEF}

Let $(\orig{\AA},\orig{\BB})$ be a two-person game in which each player
  has $n$ strategies.
Below we assume $\orig{\AA}\in \Reals{n\times n}_{[1,2]}$
  and $\orig{\BB} \in \Reals{n\times n}_{[1,2]}$.
We use the reduction introduced by
  Codenotti, Saberi, Varadarajan, and Ye \cite{CSVY}
  to map a bimatrix game to a Leontief economy.
This reduction constructs a Leontief economy with
  $(\orig{\EE},\orig{\DD}) = \pi(\orig{\AA},\orig{\BB})$ where
  the endowment matrix is simply $\orig{\EE} = \II_{2n}$,
  the $(2n)\times (2n)$ identity  matrix and
the utility matrix is given by
\[\orig{\DD}=\left(\begin{array}{ccc}0&\quad&\bar{\AA}\\\bar{\BB}&\quad&0\end{array}\right).\]

$(\orig{\EE},\orig{\DD})$ is a
   special form of Leontief exchange economies
   \cite{YeRationality,CSVY}.
It has $2n$ goods and $2n$ traders.
The $j^{th}$ trader comes to the market with one unit
  of the $j^{th}$-good.
In addition, the traders are divided into two groups
  $\calM = \setof{1,2,...,n}$ and $\calN = \setof{n+1,...,2n}$.
Traders in $\calM$ only interests in the goods associated with
  traders in $\calN$ and vice versa.

Codenotti {\em et al} \cite{CSVY} prove that
  there is a one-to-one correspondence between
  Nash equilibria of the two person game $(\AA,\BB)$ and market equilibria of Leontief economy $(\orig{\EE},\orig{\DD})$.
It thus follows from the theorem of Nash
\cite{Nash,NashNonCooperative}, that
  the Leontief economy $(\orig{\EE},\orig{\DD})$ has at least
  one equilibrium.

\newpage
We will prove the following extension of their result in the next section.

\begin{lemma}[Approximation of Games and Markets]\label{lem:app}
For any bimatrix game $(\orig{\AA},\orig{\BB})$, let $(\orig{\EE},\orig{\DD})
  = \pi(\orig{\AA},\orig{\BB})$.
Let $(\uu,\ww)$ be an $\epsilon$-approximate equilibrium
  of $(\orig{\EE},\orig{\DD})$ and assume
   $\uu=(\xx^{\top},\yy^{\top})^{\top}$ and
  $\ww=(\pp^{\top},\qq^{\top})^{\top}$.
Then, $(\xx,\yy)$ is an $O\left(n\sqrt{\epsilon}\right)$-relatively-approximate
Nash equilibrium
  for $(\orig{\AA},\orig{\BB})$.
\end{lemma}

Lemma \ref{lem:app} enables us to prove
  one of the  main results of this paper.

\begin{theorem}[Market Approximation is Likely Hard]\label{ppadb}
The problem of finding a $1/n^{\Theta(1)}$-approximate equilibrium
  of a Leontief economy with $n$ goods and $n$ traders is
\textbf{\emph{PPAD}}-hard.

Therefore, if \textbf{\emph{PPAD}} is not in \textbf{\emph{P}}, then
   there is no algorithm
   for finding an $\epsilon$-equilibrium of Leontief economies
   in time polynomial in $n$ and $1/\epsilon$.
\end{theorem}

\begin{proof}
Apply Lemma \ref{lem:app} with $\epsilon = n^{-h}$ for a sufficiently
  large constant $h$ and Corollary \ref{cor:cor}.
\end{proof}

\subsection{The Smoothed Complexity of Market Equilibria}

In the smoothed analysis of the Leontief market exchange problem,
  we assume that entries of the endowment and utility
  matrices is subject to slight random perturbations.

Consider an economy with $\left(\orig{\EE} \in\Reals{n\times n}_{[0,2]},
 \orig{\DD}\in \Reals{n\times n}_{[0,1]}\right)$.
For a $\sigma > 0$, a perturbed economy
  is defined by a pair of random matrices $\left(\DDelta^E,\DDelta^D\right)$
  where $\delta^E_{i,j}$ and $\delta^D_{i,j}$
  are independent random variables of magnitude $\sigma$.
The common two perturbation models
   are the uniform perturbation and Gaussian perturbation.
In the {\em uniform perturbation} with magnitude $\sigma$,
  a random variable is chosen uniformly from the
  interval $[-\sigma,\sigma]$.
In the {\em Gaussian perturbation} with variance $\sigma^2$,
   a random variable $\delta$ is chosen with density
\begin{equation*}
\frac{1}{\sqrt{2 \pi} \sigma} e^{-\delta^{2}/ 2
\sigma^{2}}.
\end{equation*}

Let $\DD = \max\left(\orig{\DD}+\DDelta^D, \mathbf{0}\right)$
  and let $\EE= \max\left(\orig{\EE}+\DDelta^E, \mathbf{0}\right)$.
Although we can re-normalize $\EE$ so that the sum of each row is equal to 1,
 we choose not to do so in favor of a simpler presentation.
The perturbed game is then given by $(\EE,\DD)$.

Following Spielman and Teng \cite{SpielmanTengSimplex},
  the smoothed complexity of
  an algorithm $W$ for the Leontief economy is defined as following:
Let $T_W(\EE,\DD)$ be the complexity of algorithm $W$ for
  solving a market economy defined by  $(\EE,\DD)$.
Then, the smoothed complexity of algorithm $W$ under perturbations
  $N_{\sigma } ()$ of magnitude $\sigma$ is
\[
\smoothed{W}{n,\sigma} = \max_{\orig{\DD} \in \Reals{n\times
n}_{[0,2]}, \orig{\EE}\in\Reals{n\times n}_{[0,1]}}
\expec{\EE\leftarrow N_{\sigma } (\orig{\EE }),\DD\leftarrow
N_{\sigma } (\orig{\DD })}{T_W(\EE,\DD)},
\]
where we use
  $\EE\leftarrow N_{\sigma } (\orig{\EE }) $
  to denote that $\EE $  is a perturbation of $\orig{\EE}$
  according to $N_{\sigma } (\orig{\EE })$.

An algorithm $W$ for computing Walrasian equilibria
   has {\em polynomial smoothed time complexity}
   if for all $0<\sigma < 1$ and
   for all positive integer $n$,
   there exist positive constants $c$, $k_1$ and $k_2$ such that
\[
\smoothed{W}{n,\sigma} \leq c\cdot n^{k_1} \sigma^{-k_2}.
\]
The Leontief exchange economy is in {\em smoothed polynomial time} if
  there exists an algorithm $W$ with polynomial smoothed time-complexity
   for computing a Walrasian equilibrium.

To relate the complexity of finding an approximate Nash equilibrium
  of two-person games with the smoothed complexity of Leontief
  economies,
  we  examine the equilibria of perturbations
  of the reduction presented in the last subsection.
In the remainder of this subsection,
  we will focus on the smoothed complexity under
  uniform perturbations with magnitude $\sigma$.
One can similarly extend the results to Gaussian perturbation
  with standard deviation $\sigma$.

Let $(\orig{\AA},\orig{\BB})$ be a two-person game in which each player
  has $n$ strategies.
Let $(\orig{\EE},\orig{\DD}) = \pi(\orig{\AA},\orig{\BB})$.
Let $\left(\DDelta^E,\DDelta^D\right)$ be a pair of perturbation matrices
   with entries drawn uniformly at random from $[-\sigma,\sigma]$.
The perturbed game is then given by
   $\EE = \max\left(\orig{\EE}+\DDelta^E, \mathbf{0}\right)$
   and $\DD = \max\left(\orig{\DD}+\DDelta^D, \mathbf{0}\right)$.

Let $\Pi_{\sigma}(\orig{\AA},\orig{\BB})$ be the set of
all $(\EE,\DD)$ that can be obtained by perturbing
  $\pi(\orig{\AA},\orig{\BB})$ with  magnitude $\sigma$.
Note that the off-diagonal entries of $\EE$ are
    between $0$ and $\sigma$, while
    the diagonal entries are between $1-\sigma$ and $1+\sigma$.

In the next section, we will prove the following lemma.
\begin{lemma}[Approximation of Games and Perturbed Markets]\label{lem:pert}
Let $(\orig{\AA},\orig{\BB})$ be a bimatrix game
 with $\orig{\AA},\orig{\BB} \in \Reals{n\times n}_{[1,2]}$.
For any $0 < \sigma < 1/ (8n)$,
let $(\EE,\DD) \in \Pi_\sigma(\orig{\AA},\orig{\BB})$.
Let $(\uu,\ww)$ be an $\epsilon$-approximate equilibrium
  of $(\EE,\DD)$ and assume
   $\uu=(\xx^{\top},\yy^{\top})^{\top}$ and
  $\ww=(\pp^{\top},\qq^{\top})^{\top}$.
Then, $(\xx,\yy)$ is an
$O(n\sqrt{\epsilon}+n^{1.5}\sqrt{\sigma})$-relatively-approximate
  Nash equilibrium  for $(\orig{\AA},\orig{\BB})$.
\end{lemma}

We now follow the scheme outlined in \cite{SpielmanTengSurvey}
  and used in \cite{ChenDengTeng}
  to use perturbations as a probabilistic polynomial
  reduction from  the approximation problem of two-person games
  to market equilibrium problem over perturbed
   Leontief economies.

\begin{lemma}[Smoothed Leontief and Approximate Nash]\label{lem:smoothed}
If the problem of computing an equilibrium of a Leontief economy
  is in  smoothed polynomial time under uniform perturbations,
  then for any $0< \epsilon' < 1$,
  there exists a randomized algorithm
  for computing an $\epsilon' $-approximate Nash equilibrium
  in expected time polynomial in $n$ and $1/\epsilon' $.
\end{lemma}
\begin{proof}
Suppose $W$ is an algorithm with polynomial smoothed complexity for
  computing a equilibrium of a Leontief economy.
Let $T_W(\EE,\DD)$ be the complexity of algorithm $W$ for solving
  the market problem defined by  $(\EE,\DD)$.
Let $N_{\sigma }()$ denotes the uniform perturbation with magnitude
  $\sigma $.
Then there exists constants  $c$, $k_{1}$ and $k_{2}$ such that
  for all $0 < \sigma  < 1$,

\[
\max_{\orig{\EE} \in \Reals{n\times n}_{[0,1]},
   \orig{\DD}\in\Reals{n\times n}_{[0,2]},
   } \expec{\EE\leftarrow
N_{\sigma } (\orig{\EE }),\DD\leftarrow N_{\sigma } (\orig{\DD
})}{T_W(\EE,\DD)} \leq c\cdot n^{k_{1}} \sigma^{-k_{2}}.
\]

Consider a bimatrix game $(\orig{\AA},\orig{\BB})$
   with $\orig{\AA},\orig{\BB} \in \Reals{n\times n}_{[1,2]}$.
For each $(\EE,\DD) \in \Pi_\sigma(\orig{\AA},\orig{\BB})$,
  by Lemma \ref{lem:pert}, by setting
  $\epsilon = 0$ and $\sigma = O (\epsilon'/n^{3})$,
  we can obtain an $\epsilon'$-approximate
  Nash equilibrium of $(\orig{\AA},\orig{\BB})$ in
  polynomial time from an equilibrium of $(\EE,\DD)$.

Now given the algorithm $W$ with polynomial smoothed time-complexity,
  we can apply the following randomized algorithm with the help of
  uniform perturbations to find an
  $\epsilon$-approximate Nash equilibrium of game $(\orig{\AA},\orig{\BB})$:

\newpage

\begin{quotation}
\noindent {\bf Algorithm} {\tt ApproximateNashFromSmoothedLeontief}
  $(\orig{\AA},\orig{\BB})$
\begin{itemize}
\item [1.] Let $(\orig{\EE},\orig{\DD}) =
\pi(\orig{\AA},\orig{\BB})$. 
\item [2.]
 Randomly choose a pair of perturbation matrices 
  $\left(\DDelta^E,\DDelta^D\right)$
  of magnitude $\sigma $.
\item [3.]
Let $\DD = \max\left(\orig{\DD}+\DDelta^D, \mathbf{0}\right)$
  and let $\EE= \max\left(\orig{\EE}+\DDelta^E, \mathbf{0}\right)$.

\item [4.] Apply algorithm $W$ to find an equilibrium $(\uu,\ww)$ of
  $(\EE,\DD)$.

\item [5.]  Apply Lemma \ref{lem:pert} to compute an
  approximate Nash equilibrium $(\xx,\yy)$ of $(\orig{\AA},\orig{\BB})$.
\end{itemize}
\end{quotation}
The expected time complexity of {\tt ApproximateNashFromSmoothedLeontief} is
   bounded from above by the smoothed complexity of $W$ when the
   magnitude perturbations is $\epsilon'/n^3$ and hence is at most
$c\cdot n^{k_{1}+3k_{2}} (\epsilon') ^{-k_{2}}.$
\end{proof}

We can use this randomized reduction to prove the second main
  result of this paper.

\begin{theorem}[Hardness of Smoothed Leontief Economies]
Unless \emph{\textbf{PPAD} $\subset$  \textbf{RP}},
  the problem of computing an equilibrium
  of a Leontief economy is not in  smoothed polynomial time, under
  uniform or Gaussian perturbations.
\end{theorem}
\begin{proof}
Setting $\epsilon' = n^{-h}$ for a sufficiently
  large constant and apply Lemma \ref{lem:smoothed}
  and Corollary \ref{cor:cor}.
\end{proof}

\section{The Approximation Analysis}

In this section, we prove Lemma \ref{lem:pert}. Let us first recall
  all the matrices that will be involved:
We start with two matrices $(\orig{\AA},\orig{\BB})$ of the bimatrix game.
We then obtain the two matrices  $(\orig{\EE},\orig{\DD}) = \pi
  (\orig{\AA},\orig{\BB})$ of the associated Leontief economy,
  where $\orig{\EE } = \II_{2n}$ and
\[\orig{\DD}=\left(\begin{array}{ccc}0&\quad&\bar{\AA}\\\bar{\BB}&\quad&0\end{array}\right).\]
We then perturb $(\orig{\EE},\orig{\DD})$ to obtain
  $(\EE ,\DD)$.
We can write $\DD$ as:
\[\DD=\left(\begin{array}{ccc}\ZZ&\quad&\AA\\\BB&\quad&\NN\end{array}\right)\]
where  for all $\forall i,j$,
$z_{ij},n_{ij} \in [0,\sigma]$ and
  $a_{i,j}-\orig{a}_{ij}, b_{i,j}-\orig{b}_{i,j}\in[-\sigma,\sigma]$,
Note also because $\orig{\AA}, \orig{\BB} \in
  \Reals{n\times n}_{[1,2]}$  and $0 < \sigma < 1$,
  $\AA$ and $\BB$ are uniform perturbations with magnitude $\sigma$
   of $\orig{\AA}$ and $\orig{\BB}$, respectively.
Moreover, $z_{i,j}$ and $n_{i,j}$ are $0$ with probability $1/2$ and
  otherwise, they are uniformly chosen from $[0,\sigma]$.

Now, let $(\uu,\ww)$ be an $\epsilon$-approximate equilibrium
  of $(\EE,\DD)$ and assume
   $\uu=(\xx^{\top},\yy^{\top})^{\top}$ and
  $\ww=(\pp^{\top},\qq^{\top})^{\top}$, where all vectors are column vectors.

By the definition of $\epsilon$-approximate market equilibrium, we
have:
\begin{equation}\label{approequcond3}
\left\{\begin{array}{rcl}
\ZZ\xx+\AA\yy&\leq&(1+\epsilon)\cdot\allone\\
\BB\xx+\NN\yy&\leq&(1+\epsilon)\cdot\allone\\
(1-\epsilon)\EE^{\top}\ww&\leq&\mbox{diag}(\uu)\DD^{\top}\ww\leq(1+\epsilon)\EE^{\top}\ww,\\
\end{array}\right.
\end{equation}
where $\mbox{diag}(\uu )$ is the diagonal
  matrix whose diagonal is $\uu $.
Since the demand functions are homogeneous with respect to the price vector $\ww$,
  we assume without loss of generality that
  $\pnorm{1}{\ww } = \pnorm{1}{\pp } + \pnorm{1}{\qq}=1$.

We will prove $(\xx,\yy)$ is an
  $O\left(n\sqrt{\epsilon}+n^{1.5}\sqrt{\sigma}\right)$-relatively-approximate
  Nash equilibrium of the two-person game $(\orig{\AA},\orig{\BB})$.
To this end, we first prove the following three properties of the approximate
  equilibrium $(\uu,\ww)$.
To simplify the presentation of our proofs, we will not
  aim at the best possible constants.
Instead,  we will make some crude approximations in our bounds to
  ensure the resulting formula are simple enough for readers.

\begin{property}[Approximate Price Symmetry]\label{pro:pricesym}
If $\pnorm{1}{\ww } = 1$, $0<\epsilon < 1/2$, and $0<\sigma < 1 /
  (2n)$, then
$$\frac{1-\epsilon-4n\sigma}{2-4n\sigma}\leq\pnorm{1}{\pp},\pnorm{1}{\qq}\leq\frac{1+\epsilon}{2-4n\sigma}.$$
\end{property}
\begin{proof}\label{}
Recall $\uu=(\xx^{\top},\yy^{\top})^{\top}$ and
  $\ww=(\pp^{\top},\qq^{\top})^{\top}$.
By~(\ref{approequcond3}) and the fact that the diagonal entries of
$\EE $ are at least $1-\sigma$, we have
\begin{eqnarray*}(1-\epsilon)(1-\sigma)\pnorm{1}{\pp}&\leq&
\allone^{\top}(\diag{\xx }\ZZ^{\top}\pp+\diag{\xx }\BB^{\top}\qq)=
(\ZZ\xx )^{\top}\pp + (\BB\xx )^{\top}\qq \\
&\leq&
  3n\sigma\pnorm{1}{\pp }+ (1+\epsilon)\pnorm{1}{\qq},
\end{eqnarray*}
where last inequality follows from $(\BB \xx ) \leq (1+\epsilon
)\allone $ (obtained from Equation (\ref{approequcond3})),
  its simple consequence $x_i\leq (1+\epsilon)/(1-\sigma)\leq 3,\forall i$
  (because entries of $\BB $ are between $1-\sigma$ and $2+\sigma$),
  and the fact that entries of $\ZZ $ are between $0$ and $\sigma $.

Applying, $\pnorm{1}{\qq}=\pnorm{1}{\ww}-\pnorm{1}{\pp}=1-\pnorm{1}{\pp}$ to
the inequality,
we have
\[
\pnorm{1}{\pp }\leq \frac{1+\epsilon }{(1-\epsilon ) (1-\sigma )+
(1+\epsilon )-3n\sigma } \leq \frac{1+\epsilon }{2 -4n\sigma}.
\]
Thus,
\[
\pnorm{1}{\qq} = 1 - \pnorm{1}{\pp }\geq \frac{1-\epsilon -4n\sigma }{2-4n\sigma }.
\]
We can similarly prove the other direction.
\end{proof}

\begin{property}[Approximate Utility Symmetry]\label{pro:utisym}
If $\pnorm{1}{\ww } = 1$, $0<\epsilon < 1/2$, and $0<\sigma < 1 /
  (8n)$, then
$$\frac{(1-\epsilon)(1-\sigma)(1-\epsilon-4n\sigma)}{(1+\epsilon)(2+2\sigma)}\leq
 \pnorm{1}{\xx },\pnorm{1}{\yy }
\leq\frac{(1+\epsilon)^2+n\sigma(1+\epsilon)(2-4n\sigma)}{(1-\sigma ) (1-\epsilon-4n\sigma)}.$$
\end{property}
\begin{proof}
By our assumption on the payoff matrices of the two-person
  games, $1\leq \orig{a}_{ij}, \orig{b}_{ij}\leq 2$, for all $1\leq
 i,j,\leq n$.
Thus,  $1-\sigma \leq {a}_{ij}, {b}_{ij}\leq 2+\sigma $. By
(\ref{approequcond3}) and the fact the diagonal entries of $\EE $
is at least $1-\sigma $, we have
\begin{eqnarray*}
x_i\geq\frac{(1-\epsilon)(1-\sigma)p_i}{\form{\bb_i}{\qq}+\form{\zz_i}{\pp}}
  \geq\frac{(1-\epsilon)(1-\sigma)p_i}{(2+\sigma)\pnorm{1}{\qq}+\sigma\pnorm{1}{\pp }}
\geq\frac{(1-\epsilon)(1-\sigma)(2-4n\sigma)p_i}{(2+2\sigma)(1+\epsilon)},
\end{eqnarray*}
where the last inequality follows from Property \ref{pro:pricesym}.
Summing it up, we obtain,
$$\pnorm{1}{\xx
}\geq\frac{(1-\epsilon)(1-\sigma)(2-4n\sigma)}{(2+2\sigma)(1+\epsilon)}
  \pnorm{1}{\pp }\geq
\frac{(1-\epsilon)(1-\sigma)(1-\epsilon-4n\sigma)}{(1+\epsilon)(2+2\sigma)},$$
where again, we use Property \ref{pro:pricesym} in the last inequality.

On the other hand, from (\ref{approequcond3}) we have
\begin{eqnarray}\xx_i
  &\leq&\frac{(1+\epsilon)\form{\ee_i}{\ww}}{\form{\bb_i}{\qq}+\form{\zz_i}{\pp}}
  \leq\frac{(1+\epsilon)(p_i+\sigma)}{\form{\bb_i}{\qq}} \quad \quad
\quad
\mbox{\hfill (using $\pnorm{1}{\ww}=1$ and the property
of $\ee_{i}$)}\nonumber\\
&\leq&\frac{(1+\epsilon)(p_i+\sigma)}{(1-\sigma )\pnorm{1}{\qq}}
  \leq\frac{(1+\epsilon)(p_i+\sigma)(2-4n\sigma)}{(1-\sigma ) (1-\epsilon-4n\sigma)}.\label{eqn:half}
\end{eqnarray}

Summing it up, we obtain,
\begin{eqnarray*}
\pnorm{1}{\xx}&\leq&\frac{(1+\epsilon)(2-4n\sigma)}{(1-\sigma )
(1-\epsilon-4n\sigma)}
  \pnorm{1}{\pp}+\frac{n\sigma(1+\epsilon)(2-4n\sigma)}{(1-\sigma) (1-\epsilon-4n\sigma)}\\
&\leq&\frac{(1+\epsilon)^2+n\sigma(1+\epsilon)(2-4n\sigma)}{(1-\sigma ) (1-\epsilon-4n\sigma)}.
\end{eqnarray*}
We can similarly prove the bound for $\pnorm{1}{\yy}$.
\end{proof}

\begin{property}[Utility Upper Bound]\label{pro:utiupper}
Let $\ss=\ZZ\xx+\AA\yy$ and $\ttt=\BB\pp+\NN\yy$. Let
  $\lambda=\max\set{\epsilon,n\sigma}$.
Under the same assumption as Property \ref{pro:utisym},
  if $s_i\leq (1+\epsilon)(1-\sigma)-\sqrt{\lambda}$,
\[
x_i\leq
\frac{(1+\epsilon)(2-4n\sigma)(5\sqrt{\lambda}+\sigma)}{(1-\sigma ) (1-\epsilon-4n\sigma)}.
\]
Similarly,
  if $t_i\leq (1+\epsilon)(1-\sigma)-\sqrt{\lambda}$, then
\[
y_i\leq
\frac{(1+\epsilon)(2-4n\sigma)(5\sqrt{\lambda}+\sigma)}{(1-\sigma ) (1-\epsilon-4n\sigma)}.
\]
\end{property}
\begin{proof}
By (\ref{approequcond3}), we have
\begin{eqnarray*}
(1-\epsilon)(1-\sigma)\pnorm{1}{\ww}&\leq&(1-\epsilon)\allone^{\top}\EE^{\top}\ww\leq
\uu^{\top}\DD^{\top}\ww = \form{\ss}{\pp}+ \form{\ttt}{\qq}\\
& = & \sum_{j\neq i}s_{j}p_{j} + \form{\ttt }{\yy } + s_{i}p_{i}\\
&\leq&(1+\epsilon)\allone^{\top}\EE^{\top}\ww-(1+\epsilon)\form{\ee_i}{\ww}+s_ip_i\\
&\leq&(1+\epsilon)(1+n\sigma)\pnorm{1}{\ww}-(1+\epsilon)(1-\sigma)p_i+s_ip_i,
\end{eqnarray*}
where the inequality immediately after the second equation
  follows from the last inequality of (\ref{approequcond3}).

Thus,
$[(1+\epsilon)(1-\sigma)-s_i]p_i\leq
(1+\epsilon)(1+n\sigma)-(1+\epsilon)(1-\sigma)\leq
2\epsilon+3n\sigma$.
Consequently, if $(1+\epsilon)(1-\sigma)-s_i\geq \sqrt{\lambda}$,
   then $p_i\leq (2\epsilon+3n\sigma)/\sqrt{\lambda }\leq 5\sqrt{\lambda}$.
By Equation (\ref{eqn:half})
\begin{eqnarray*}
x_i\leq \frac{(1+\epsilon)(2-4n\sigma)(p_i+\sigma)}{(1-\sigma ) (1-\epsilon-4n\sigma)}
\leq\frac{(1+\epsilon)(2-4n\sigma)(5\sqrt{\lambda}+\sigma)}{(1-\sigma ) (1-\epsilon-4n\sigma)}.
\end{eqnarray*}
We can similarly establish the bound for $y_{i}$.
\end{proof}

We now use these three properties to prove Lemma \ref{lem:pert}.


\begin{proof}{\bf [of Lemma \ref{lem:pert}]}
In order to prove that $(\xx,\yy)$ is a $\delta$-relatively approximate Nash
  equilibrium for $(\orig{\AA},\orig{\BB})$,
  it is sufficient to establish:
\begin{equation}\label{approxnashcond}
\left\{\begin{array}{l}
\xx^{\top}\orig{\AA }\yy\geq(1-\delta)
  \max\limits_{\pnorm{1}{\tilde{\xx}}=\pnorm{1}{\xx}}\tilde{\xx}^{\top}\orig{\AA}\yy\\
\yy^{\top}\orig{\BB}^{\top}\xx\geq(1-\delta)
\max\limits_{\pnorm{1}{\tilde{\yy}}=\pnorm{1}{\yy}}\tilde{\yy}^{\top}\orig{\BB}^{\top}\xx.\\
\end{array}\right.
\end{equation}

Let $\ss=\ZZ\xx+\AA\yy$.
We observe,
\begin{eqnarray*}
\xx^{\top}\orig{\AA}\yy&=&\xx^{\top}\left(\AA\yy+\ZZ\xx-\ZZ\xx+(\orig{\AA}-\AA)\yy\right)
  =\xx^{\top}\ss-\xx^{\top}\left(\ZZ\xx+\left(\AA-\orig{\AA}\right)\yy\right)\\
&\geq&\xx^{\top}\ss-\pnorm{1}{\xx}\left(\pnorm{\infty}{\ZZ\xx}
  +\pnorm{\infty}{(\AA-\orig{\AA})\yy}\right)\geq
\xx^{\top}\ss-\sigma\pnorm{1}{\xx}\left(\pnorm{1}{\xx}+\pnorm{1}{\yy}\right)\\
&\geq&\xx^{\top}\ss-\frac{2\sigma(1+\epsilon)^2+2n\sigma^2(1+\epsilon)(2-4n\sigma)}
{(1-\sigma ) (1-\epsilon-4n\sigma)}\pnorm{1}{\xx}\\
&=&\xx^{\top}\ss-O(\sigma)\pnorm{1}{\xx},
\end{eqnarray*}
where the last inequality follows from Property \ref{pro:utisym}.

Let $\lambda=\max (\epsilon,n\sigma)$.
By Property \ref{pro:utiupper}, we can estimate the lower bound of $\form{\xx}{\ss}$:
\begin{eqnarray*}\form{\xx}{\ss}&=&\sum\limits_{i=1}^nx_is_i
\geq\sum\limits_{i:s_i>(1+\epsilon)(1-\sigma)-\sqrt{\lambda}}x_is_i\\
&\geq&\left[(1+\epsilon)(1-\sigma)-\sqrt{\lambda}\right]
\left(\pnorm{1}{\xx}-n\frac{(1+\epsilon)(2-4n\sigma)(5\sqrt{\lambda}+\sigma)}
  {(1-\sigma ) (1-\epsilon-4n\sigma)}\right)\\
&\geq&\pnorm{1}{\xx}\left(1-O \left( \sqrt{\lambda}\right)\right)
  \left(1-n\frac{(1+\epsilon)(2-4n\sigma)(5\sqrt{\lambda}+\sigma)}
  {(1-\sigma ) (1-\epsilon-4n\sigma)\pnorm{1}{\xx}}\right)\\
&=&\pnorm{1}{\xx}\left[1 - O\left(n\sqrt{\lambda}+n\sigma\right)\right].
\end{eqnarray*}

On the other hand, by (\ref{approequcond3}), we have
  $\AA \yy \leq (1+\epsilon)\allone $ and hence
\[\max\limits_{\pnorm{1}{\tilde{\xx}}=\pnorm{1}{\xx}}\tilde{\xx}^{\top}\bar{\AA}\yy
= \pnorm{1}{\xx}\pnorm{\infty}{\orig{\AA} \yy}
  \leq(1+\epsilon+\sigma\pnorm{1}{\yy})\pnorm{1}{\xx}\leq(1+\epsilon+O(\sigma))\pnorm{1}{\xx}.\]

Therefore,
\begin{eqnarray*}
\xx^{\top}\orig{\AA}\yy&\geq&\xx^{\top}\ss-O(\sigma)\pnorm{1}{\xx}\\
&\geq&\|\xx\|_1\left[(1-O\left(n\sqrt{\lambda} + n\sigma\right)-O(\sigma)\right]\\
&\geq&\frac{1}{1+\epsilon+O(\sigma)}\left[1-O\left(n\sqrt{\lambda}
+ n\sigma\right) \right]
 \max\limits_{\pnorm{1}{\tilde{\xx}}=\pnorm{1}{\xx}}\tilde{\xx}^{\top}\orig{\AA}\yy\\
&=&\left(1-O\left (n\sqrt{\lambda}+n\sigma\right)\right)
  \max\limits_{\pnorm{1}{\tilde{\xx}}=\pnorm{1}{\xx}}\tilde{\xx}^{\top}\orig{\AA}\yy.
\end{eqnarray*}
We can similarly prove
\begin{eqnarray*}
\yy^{\top}\orig{\BB}^{\top}\xx =\left(1-O\left
(n\sqrt{\lambda}+n\sigma\right)\right)
  \max\limits_{\pnorm{1}{\tilde{\yy}}=\pnorm{1}{\yy }}\tilde{\yy}^{\top}\orig{\BB}^{\top}{\xx}.
\end{eqnarray*}
We then use the inequalities
$\sqrt{\lambda} =\sqrt{\max (\epsilon,n\sigma )}\leq\sqrt{ \epsilon
+n\sigma}\leq
\sqrt{\epsilon} +\sqrt{n\sigma}$ and $\sigma \leq \sqrt{\sigma }$ to complete the proof.

\end{proof}
\section{Remarks and Open Questions}

In this section, we briefly summarize some remarkable
   algorithmic accomplishments
  in the computation of (approximate) market equilibria obtained
  prior to this work.
We then present some open questions motivated by these and our
  new results.

\subsection{General Market Exchange Problems and Algorithmic Results}

In our paper, we have focused on Leontief market exchange economies.
Various other market economies have been considered in the literature
  \cite{ScarfBook,DengHuangSurvey}.

An instance of a general market exchange economy 
  with $n$ traders and $m$ goods
  is given by the endowment matrix $\EE$ together with $n$ utilities
  functions $\vs{v}{1}{n}$.
Then an {\em equilibrium price vector} is a vector $\pp$ satisfying
\[
\exists\ \XX \in \Reals{m\times n}, \XX\allone \leq \EE\allone,\
\XX^{\top} \pp \leq  \EE^{\top}\pp, \ \mbox{ and } u_j(\xx_j) =
\max\setof{u_j(\xx_j'): \form{\xx_j'}{\pp} \leq
\form{\ee_j}{\pp}}.
\]
The pairs $(\XX,\pp)$ and
  $\left(\left[u_1(\xx_1),...,u_n(\xx_n)\right],\pp\right)$ is
  as also referred to as an equilibrium of the exchange market
  $\left(\EE,(\vs{u}{1}{n})\right)$.

The celebrated theorem of Arrow and Debreu \cite{ArrowDebreu}
  states that if all utility functions are concave,
  then the exchange economy has an equilibrium.
Moreover, if these functions are strictly concave, then
  for each equilibrium price vector $\pp$, its associated
  exchange $\XX$ or utilities is unique.





A popular family of utility functions is the
  CES (standing for Constant Elasticity of Substitution)
  utility functions.
It is specified by an $m\times n$ demand matrix $\DD$.
The utility functions are then defined with the help of
  an additional parameter $\rho\in (-\infty,1]\setminus \setof{0}$:
\[u^{(\rho)}_j(\xx_j)=\left(\sum\limits_{i=1}^m d_{ij}x_{ij}^{\rho}\right)^{\frac{1}{\rho}}\]

As $\rho \rightarrow -\infty$, CES utilities become the Leontief utilities.
When $\rho = 1$, the utility functions are linear functions.

Remarkablely, an (approximate) equilibrium of an exchange economy
  with linear utilities functions can be found in polynomial time
  \cite{NenakhovPrimak,Jain,YeArrow,JainMahdianSaberi}.
In fact, Ye shows that
  an $\epsilon$-equilibrium of such an market with $n$ traders and
  $n$ goods can be found in $O(n^4\log (1/\epsilon))$ time \cite{YeArrow}.
If data is given as rational numbers of $L$-bits, then an exact
  equilibrium can be found in $O(n^4L)$ time.

A closely related market exchange model is the Fisher's model \cite{ScarfBook}.
In this model, there are two types of traders: {\em producers} and
  {\em consumers}.
Each consumer comes to the market with a budget and
  a utility function.
Each producer comes to the market with an endowment of goods and
 will sell them to the consumers for money.
An equilibrium is a price vector $\pp$ for goods so that
  if each consumer spends all her budget to maximize
  her utilities, then the market clears, i.e., at
  the end of the exchange, all producers  sold out.

Even more remarkablely, an (approximate)
  equilibrium in a Fisher's economy with
  any CES utilities can be found in polynomial time
  \cite{EisenbergGale,YeArrow,YeRationality,Devanur,JainVaziraniYe}.

\subsection{Open Questions}

Our results as well as the combination of
  Codenotti, Saberi, Varadarajan, and Ye \cite{CSVY} and
  Chen and Deng \cite{ChenDeng} demonstrate that
  exchange economies with Leontief utility functions
  are fundamentally different from economies with
  linear utility functions.
In Leontief economies, not only finding an exact equilibrium
  is likely hard, but finding an approximate
  equilibrium is just as hard.

Although, we prove that the computation of an
   $O(1/n^{\Theta(1)})$-approximate
   equilibrium of Leontief economies is \textbf{PPAD}-hard.
 our hardness result does not cover
   the case when $\epsilon$ is a constant
   between $0$ and $1$.
The following are two optimistic conjectures.

\begin{conjecture}[PTAS Approximate LEONTIEF]
There is an algorithm to find an $\epsilon$-approximate
  equilibrium of a Leontief economy in
  time $O(n^{k+\epsilon^{-c}})$ for some positive constants $c$ and $k$.
\end{conjecture}

\begin{conjecture}[Smoothed LEONTIEF: Constant Perturbations]
There is an algorithm to find an equilibrium of a Leontief economy
  with smoothed time complexity $O(n^{k+\sigma^{-c}})$
  under perturbations with magnitude $\sigma$,
  for some positive constants $c$ and $k$.
\end{conjecture}

\section{Acknowledgments}
We, especially the second author, would like to thank Xi Chen and 
  Xiaotie Deng for their invaluable impact to this paper.
It was their remarkable breakthrough
  result that 2-NASH is \textbf{PPAD}-complete
  and their joint work with the second author on the complexity
  of approximate equilibria of 2-NASH that made this paper possible.

\bibliographystyle{plain}

\bibliography{LEONTIEF}
\end{document}